# STUDIES OF POLYMER DEFORMATION AND RECOVERY IN HOT EMBOSSING

*X. C. Shan[1]\*,   Y. C. Liu[1], H. J. Lu[1], Z. F. Wang[1] and Y. C. Lam[2]*

[1] Singapore Institute of Manufacturing Technology (SIMTech),
71 Nanyang Drive, Singapore 638075
[2] Nanyang Technological University, 50 Nanyang Ave, Singapore 639798
\*E-mail (X.C. Shan): xcshan@simtech.A-star.edu.sg

## ABSTRACT

In large area micro hot embossing, process temperature plays a critical role in both the local fidelity of microstructure formation and global uniformity. Micro embossing at the lowest temperature with acceptable fidelity can improve the global flatness after demolding. This paper focuses on polymer deformation and recovery in micro embossing when the process temperatures are below its glass transition temperature ($T_g$). In this study, PMMA (Polymethyl Methacrylate) substrates ($T_g$ =105 °C) were employed as the process material, and the process temperature ranged from 25 °C to $T_g$. It was found that at temperature below $T_g$-55 °C, significant recovery occurred after processing, but it was still possible to form permanent structures with sufficiently high loading stress. With an increase in temperature, plastic deformation increased and was the dominate polymer deformation for permanent cavities formation. However, the formation of protrusive structures was not complete since there was little polymer flow. The polymer will lose its storage modulus at a raised temperature and microstructures could be formed with high fidelity. A compromise between the local fidelity and global flatness has to be reached in micro hot embossing.

## 1. INTRODUCTION

An amorphous polymer such as PMMA (Polymethyl Methacrylate) behaves as a viscoelecsic solid at raised temperature. At a temperature far above its $T_g$, the polymer behaves more like a viscous liquid and flows relatively easily into the corners of an embossing mold. As such, microstructures with high fidelity can be obtained [1, 2]. In large area hot embossing, the process temperature, which has a direct effect on the visco-elastic behavior of a polymer, plays a critical role in both the local fidelity and global uniformity of microstructure formation. To improve the fidelity of embossed microstructures, high process temperature (as high as $T_g$ +60°C) are preferred. High temperature embossing, however, has its disadvantages, such as difficulty in demolding and significant residual thermal stress due to the different coefficients of thermal expansion between the mold and the polymer. As a result, the embossed devices will suffer from global warpage or distortion due to mechanical and residual thermal stresses [3]. However, with low temperature hot embossing, the global flatness of an embossed substrate could well be improved significantly. Hence, it is important to perform micro embossing at a process temperature as low as possible, but with acceptable fidelity, especially for large area and high aspect embossing.

This paper focuses on the microstructure formation, polymer deformation and recovery in micro hot embossing at temperature lower than a polymer's $T_g$. Both an indentation system [4] and a micro embosser were used to investigate the relationship of microstructure formation versus process temperature and applied stress.

Indentations conducted from 25°C to 50°C ($T_g$ - 55°C) indicated that PMMA behaved like a visco-elastic solid, with viscoelesticity dominated the formation of microstructures with large recovery after demolding. From 85°C ($T_g$ -20 °C) to near $T_g$, plastic deformation dominated the structure formation, and permanent cavities could be formed on PMMA substrates with decreased recovery with an increase in temperature. However, profile formation of protrusive structures over this temperature range was not complete since there was little polymer flow. It is also found that plastic deformation still contributed significantly in micro embossing even above $T_g$. With increasing process temperature, microstructures could be formed under lower loading. Considering the local fidelity and global flatness of embossed substrates, micro hot embossing at low process temperature, but with acceptable fidelity, should be preferred.

## 2. MECHANICS OF POLYMER DEFORMATION

### 2.1. Viscoelasticity of Polymers

An amorphous polymer like PMMA exhibits highly temperature-dependent viscoelasticity, which determines the elastic and permanent deformation of the polymer. Viscoelasticity, which is the response of the polymer to an applied stress, contains both an elastic and a viscous components [5]. The total deformation of the polymer under a mechanical stress will contain both permanent deformation as well as recoverable elastic deformation on the removal of the applied stress.

The total deformation can be figuratively considered as consisting of: (1) instantaneous elastic deformation; (2) retarded elastic deformation; and (3) viscous deformation. Both elastic deformations are recoverable after unloading, the retarded elastic and viscous deformation, however, are highly temperature-dependent components [5]. A viscoelastic polymer will also exhibit creep, which is a time- and temperature- dependent deformation when the applied stress is kept constant. Creep includes both the retarded elastic and viscous components.

## 2.2. Nano-indentation to Simulate Hot Embossing

A nano-indentation system (Micro Materials Ltd) was employed to investigate precisely the relationship of microstructure formation versus applied stress at a certain temperature [6]. The PMMA sample was mounted onto a hot stage using high temperature adhesive. A Berkovich diamond indenter was mounted onto a pendulum. A varying load was applied by means of a coil and magnet located at the top of the pendulum and the indenter was impressed into the sample surface. The resultant indent depth into the PMMA surface was measured by a capacitive transducer and displayed as a function of load. A thermal shield was interposed between the loading head and the hot stage to prevent any thermal interference, and glass wools were used to isolate the diamond indenter-specimen from the environment to eliminate heat flow.

## 2.3. Nano-indentation of PMMA

Experiments were carried out using 1 mm thick PMMA substrates. The indented depth was controlled and the applied force at a certain indent depth was measured. The process temperature investigated ranged from 25 °C to 85 °C. The microstructures formed by indentation included a single micro cavity and an array of micro cavities.

Fig. 1 shows the curves of applied loading versus indent depth at 25 °C and 85 °C obtained via the indentation system. The depth of indentation was set to 3μm with the same loading rate of 1 mN/second. The loading curve, creep (hold) and unloading curve were AB', B'C' and C'D' respectively at 85 °C, and AB, BC and CD respectively at 25 °C. The loading force required for indenting a 3μm depth at 85 °C was 24 mN (equivalent to 113 MPa), while the loading force was 45 mN (equivalent to 210 MPa) at 25 °C. Since the loading period lasted for more than 20 seconds, it was estimated that there was also creep component in the loading curves. There was instantaneous recovery during unloading, and the indent depth after unloading was 2.34μm for indentation at 85°C and 2.25μm for 25°C. Indentation performed at 50 °C demonstrated similar phenomena.

Fig. 2 shows the creep behavior when the applied loading was kept constant. The measurement of creep was conducted using the indentation system under different temperature and loading stress. The creep curves at indent depth of 3μm at 85 °C (curve-1) and 25 °C (curve-2) were obtained directly from Fig.1, and those at indent depth of 1 μm (curve-3 for 85 °C) and curve-4 for 25 °C) were from another experiment. The slope of the creep curves, which represents the creep rate, decreased with the progress of time, but did not approach zero; this means that the polymer would exhibit further creep if the measurement was continued. Fig.2 also demonstrates that the creep behavior observed was a time-, temperature- and stress-dependent deformations.

Fig. 3 shows the loading-hold-unloading versus indent depth at 85°C (Tg-20°C). Holding force was kept constant when the indent depth reached 1.1μm, 2.2μm and 3.0μm, respectively. Both the loading and unloading rates were set to 1 mN/second, and the creep time at hold was set to 18 seconds. It was observed that the creep behavior was a time- and stress- dependent deformations. The resultant indent depth was 0.77μm, 1.56μm and 2.34μm, respectively after unloading. In these three cases, the instantaneous elastic recovery was proportional to its indent depth at a given temperature.

In addition to the instantaneous recovery after the removal of the applied stress, it was found that the structures also exhibited retarded or long term recovery. Fig. 4 shows the AFM image of the cavity indented with a load of 24 mN (equivalent to 113 MPa) at 85 °C. As shown by curve-3 in Fig. 3, the feature had a depth of 2.34μm after unloading. However, AFM measurement at room temperature several hours later shows that the depth was about 1.38 μm. This indicates the occurrence of retarded recovery while the PMMA substrate was cooled and detached from the system. This retarded recovery would last for a long duration, and the stable depth measured one week after indentation was 1.25 μm. Table 1 illustrates depths measured under different temperatures and durations.

The results in Figs. 1-4 indicate that at temperature lower than its Tg, an amorphous polymer subjected to an applied stress exhibits viscoelastic deformation with both instantaneous and retarded recovery. A permanent deformation, however, can also be obtained. This indicates that micro embossing can be performed at a lower temperature than the polymer's Tg.

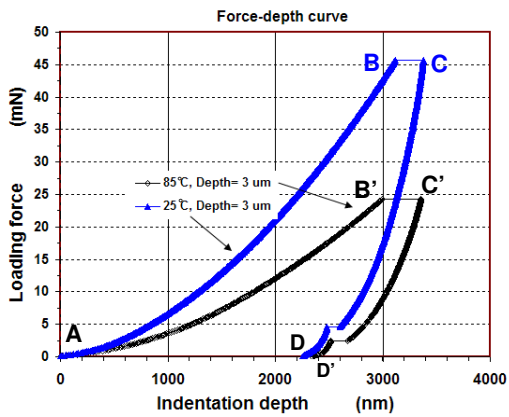

Fig.1  Loading-hold-unloading versus indent depth on PMMA substrates at 25°C and 85 °C.

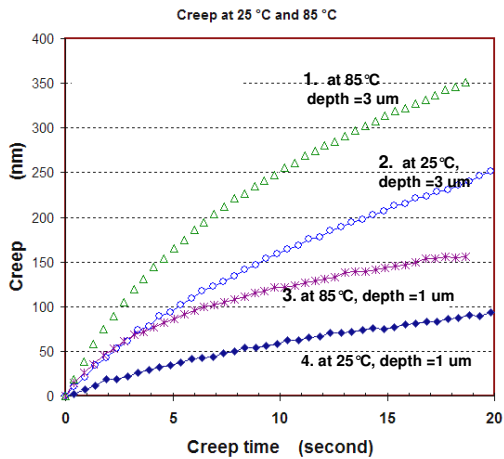

Fig. 2 Creep (hold) versus time with a constant load at 25°C and 85 °C

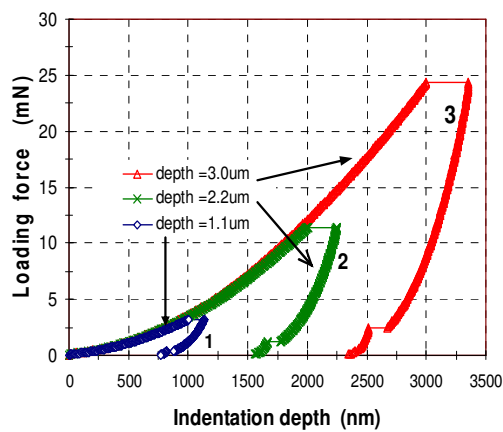

Fig.3 Loading-hold-unloading versus indent depth at 85 °C under different loading force. The creep time during hold for these three cases was 18 seconds.

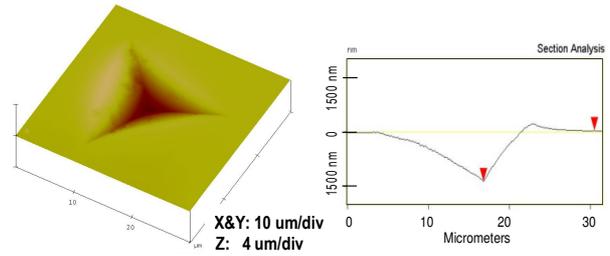

Fig. 4 AFM image of an indented cavity and its cross-section. The cavity was impressed to 3μm deep at 85 °C; but the AFM image was obtained at room temperature.

Table 1: Depth of indented cavities under different temperatures and durations

| Temperature (°C) | Original Depth (μm) | Depth aft creep (μm) | Depth after unloading (μm) | AFM image (μm) | Depth 1-week after (μm) |
|---|---|---|---|---|---|
| 25°C | 1.04 | 1.13 | 0.72 | 0.37 | 0.34 |
|  | 3.12 | 3.37 | 2.25 | 1.07 | 0.99 |
| 85°C | 1.0 | 1.15 | 0.77 | 0.48 | 0.46 |
|  | 3.0 | 3.35 | 2.34 | 1.38 | 1.26 |

## 3. MICRO HOT EMBOSSING

A desk-top embosser, with a hydraulic press mechanism for force loading, was used for microstructure formation. The system has a loading capacity of 15 tons. The heaters, attached on both upper and bottom fixtures, can be heated up to 250 °C. Silicon molds etched via DRIE were used in this investigation.

The hot embossing process was carried out by using a silicon mold with micro pillar arrays, the size and height of the pilliars were 50μm × 50μm and 7μm, respectively. A stylus profilometer and a Wyko® interferometer were used to evaluate the profiles of embossed structures under different process parameters, which included temperature, applied force and pressing time. The process temperature ranged from $T_g$ -20°C to $T_g$ ($T_g$ =105 °C for PMMA). The embossing load was set to 100 kgf, which is eqivenlent to an average pressure of 25 MPa.

It was found that stable cavity depth could not be obtained at 85°C ($T_g$ -20°C). This is because the applied force was much lower than that described in Fig. 3. It was also found that permanent and stable cavities could be formed at 90 °C and above, and the fidelity increased with an increase in the process temperature and applied force. Fig. 5 illustrates the profiles of the embossed cavitity arrays at different temperatures measured by a stylus profilometer, with the distance between the two cavities being 150 μm. Considering that polymer flow was extremly restricted at a temperature below its Tg, it was believed that the formation of microstructures in Fig. 5 was predominantly attributed to plastic deformation.

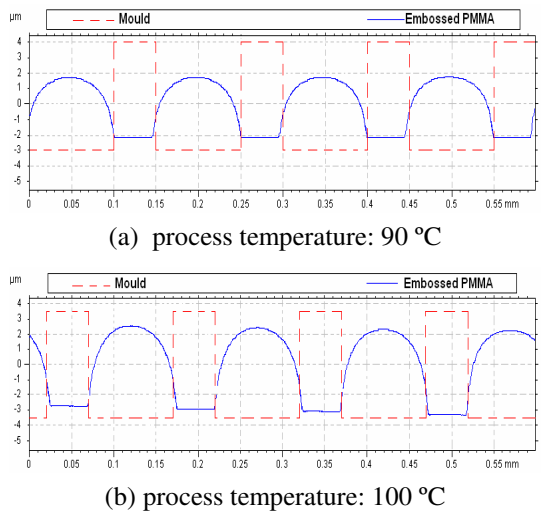

(a)  process temperature: 90 ℃

(b) process temperature: 100 ℃

Fig. 5 Cross section of cavities embossed at 100 kgf. (25 MPa; pillar size: 50μm × 50μm)

Fig. 6 shows the surface profile measured by a Wyko® interferometer. Embossing was performed at 100 ˚C using a mold with pillar size of 50 μm × 50 μm. The topography of the embossed PMMA surface indicated that micro cavities were formed with a dominance of plastic deformation. The cross section in X-X and Y-Y sections revealed the wavy surface of the protusive structures adjacent to the cavities. The highest points of the wavy surface in X-X section were located at the furthest locations from the cavities; while the lowest points were close to the cavities.

The wavy surfaces along the X-X and Y-Y sections were formed due to plastic deformation during hot embossing. This was due to the fact that, in the experiments, limited viscous flow at a temperature below Tg prevented the polymer to move along the vertical sidewalls of the mold. Further experimental studies showed that the complete structure could be replicated at a temperature above 110 ℃ (Tg+5˚C) with a loading pressure of about 25 MPa.

## 4.  SUMMARY

Polymer deformation and recovery in microstructure formation were studied using nano-indentation technique. When the temperature was below its Tg, an amorphous polymer exhibited viscoelastic deformation under an applied stress with both instantaneous and retarded recovery. Permanent deformation, however, could also be obtained after recovery. This indicates that hot embossing can be performed at a lower temperature than the polymer's Tg. Micro hot embossing using PMMA were investigated when the process temperature was below or near its Tg. It is found that for temperatures of Tg-20 °C and above, permanent cavities were formed on 1mm thick PMMA substrate without obvious retarded recovery, although plastic deformation dominated the structure formation. With an increase in process temperature, microstructures could be imprinted under a lower loading force. A compromise between the fidelity of micro-structures and global flatness of embossed substrates has to be considered for the optimization of the process parameters. Micro hot embossing at low process temperature, but with acceptable fidelity, should be preferred.

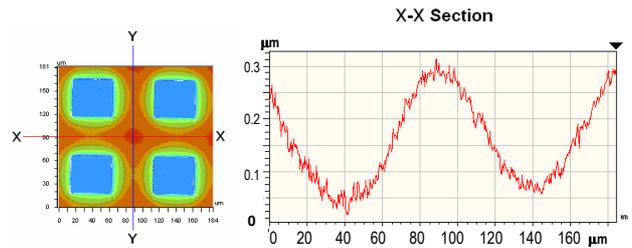

Fig. 6 The wavy surface topography from a Wyko interferometer analysis (100 ℃, 25 MPa)